\documentstyle[prd,preprint,tighten,aps,eqsecnum,%
amssymb,newlfont]{revtex}
\newcommand{\be}{\begin{equation}}
\newcommand{\ee}{\end{equation}}
\newcommand{\ba}{\begin{eqnarray}}
\newcommand{\ea}{\end{eqnarray}}
\newcommand{\no}{\nonumber}
\begin{document}
\draft
\preprint{
\begin{tabular}{r}
UWThPh-1998-40\\
PRL-TH-1998\\
September 1998
\end{tabular}
}
\title{The field-theoretical approach to coherence\\ 
in neutrino oscillations}
\author{W. Grimus and P. Stockinger}
\address{Institute for Theoretical Physics, University of Vienna\\
Boltzmanngasse 5, A-1090, Vienna, Austria}
\author{S. Mohanty}
\address{Theory Group, Physical Research Laboratory\\
Ahmedabad - 380 009, India}
\maketitle
\begin{abstract}
We study the conditions for the existence of neutrino oscillations in
the field-theoretical approach which combines neutrino production
and detection processes in a single Feynman graph. The
``oscillating neutrino'' is represented by an inner line
of this graph where, due to the
macroscopic distance $L$ between source and detector, 
the neutrino propagators for neutrinos with definite mass
are replaced by the projection operators unto the
neutrino states on mass shell. We use as a concrete model
reaction the neutrino source and detector as given in the LSND
experiment and we carefully take into account the finite
lifetime of the stopped muons which provide the $\bar\nu_\mu$ beam.
We show that the field-theoretical approach provides a
solid method to locate all possible conditions and allows 
to separate unambiguously their different origins. Some of these conditions
are independent of $L$ whereas others state that
coherence is lost when $L$ exceeds a certain
``coherence length''. Also it turns out that, 
at least in the concrete situation considered here,
the concept of neutrino wave packets is not supported by the
field-theoretical approach for realistic experimental conditions,
i.e., the neutrino energy spread is incoherent in origin.
\end{abstract}


\narrowtext
\section{Introduction}
\label{introduction}

The standard treatment of neutrino oscillations \cite{pon,kim} 
provides a beautiful and simple picture of this important
phenomenon. With the mixing matrix relating the left-handed neutrino 
flavor fields with the left-handed neutrino mass eigenfields defined by
$\nu_{L \alpha} = \sum_j U_{\alpha j} \nu_{Lj}$
($\alpha = e, \mu, \tau, \ldots$)
it allows to derive the oscillation probabilities for antineutrinos
\ba
P_{\bar\nu_\alpha\to\bar\nu_\beta} & = &
\left|
\sum_j U^*_{\beta j} U_{\alpha j} 
\exp \left( -i \frac{m^2_j L}{2E_\nu} \right)
\right|^2 \no\\
& = & \sum_j |U_{\beta j}|^2 |U_{\alpha j}|^2 +
2\, \mbox{Re} \left\{ \sum_{j>k} 
U^*_{\beta j} U_{\alpha j} U_{\beta k} U^*_{\alpha k}
\exp \left( -i \frac{\Delta m^2_{jk} L}{2E_\nu} \right) \right\} \,,
\label{P}
\ea
valid in the ultrarelativistic limit
with $\Delta m^2_{jk} \equiv m^2_j-m^2_k$ where 
$m_1 \leq m_2 \leq \ldots$ denote the neutrino masses, $L$ is the
distance between neutrino source and detector and $E_\nu$ is the
neutrino energy.
The probability for neutrinos is obtained from Eq.(\ref{P}) by the
substitution $U \to U^*$. In the following, Greek indices always indicate 
neutrino flavors and Latin indices mass eigenstates or fields.
However, after a closer look one discovers that the standard derivation
of Eq.(\ref{P}) needs clarification in several points (see, e.g., 
Ref.\cite{rich} for a summary of these problems). This has first been 
attempted by using neutrino wave packets 
\cite{N76,kayser,GKL,KNW96,lowe,ancochea,grossman,GKL97,GK97}, 
whereas Ref.\cite{rich}
has pioneered the idea of considering the complete neutrino production --
detection chain using only those quantities for the description of
neutrino oscillations which are really observed
or manipulated in oscillation experiments \cite{GS96,cam,KW97}
in order to obtain unambiguous results. See also Ref.\cite{GKLL} for a
sort of combined field theory -- wave packet approach.
The present interest in theoretical treatments of neutrino oscillations
can be phrased by the following question:
\emph{Under which conditions is formula (\ref{P}) valid?}
Since the neutrino wave packet formalism does not work with the physical
observables we find the field-theoretical approach treating neutrino
production and detection \cite{GS96} the most appropriate, 
unambiguous and general way to analyse the problem of coherence in neutrino
oscillations. In particular, when several quantities defining
a length are involved, an improvement of the wave packet approach is
called for in order to distinguish the roles and origins of these lengths.

The transition probability (\ref{P}) is given by the square of the sum
over the amplitudes of the neutrino mass eigenstates, i.e., by a
coherent summation over the mass eigenstates. The first term in the
second line of Eq.(\ref{P}) represents the purely incoherent summation
over the mass eigenstates whereas the second term denotes the
interference terms. The exponentials 
$\exp (-i2\pi L/L_{jk}^{\mathrm{osc}})$ 
with the oscillation lengths defined by
\be
L_{jk}^{\mathrm{osc}} \equiv \frac{4\pi E_\nu}{\Delta m^2_{jk}} 
\ee
show the oscillatory behaviour of the transition probability as a
function of $L/E_\nu$. Eq.(\ref{P}) is a theoretical expression
without regard to an actual experimental situation.
In the description of a neutrino oscillation
experiment it is possible that, after taking into account the
experimental conditions, some or all of the interference terms drop
out as a consequence of certain averaging or suppression mechanisms
to be discussed in the following. Note that the effect of such
mechanisms is equivalent to a partial or complete incoherent summation
over the neutrino mass eigenstates.

One such class of mechanisms is given by all effects 
leading to an energy spread 
of the neutrino beam.  It has been shown in Ref.\cite{KNW96} that
if we label such effects by $a$ then each of these effects giving
an energy spread $\Delta E_a$ leads to a coherence length
\be\label{Lcoh}
L^{\mathrm{coh}}_{a;\,jk} = 
L^{\mathrm{osc}}_{jk} \frac{E_\nu}{\Delta E_a}
\ee
independent of the fact whether this spread has to be interpreted as a
coherent or incoherent effect. In the context of a neutrino energy
spread ``incoherent'' 
means that single neutrinos have a definite energy but the neutrino
beam has an energy spread whereas by ``coherent'' 
it is understood that a single neutrino state is a superposition of different
energies.\footnote{The notion of a neutrino wave packet is synonymous
with the presence of a coherent neutrino energy spread.}
Note, however, that in the field-theoretical
approach the notions ``incoherent'' and ``coherent'' energy spread have 
well-established and precise definitions:
``incoherent'' means that the summation over different neutrino
energies happens in the cross section of the total production --
detection process whereas a summation over different neutrino energies
in the amplitude is called a ``coherent'' energy
spread.\footnote{Hence, whether the summation over neutrino mass
eigenstates or the summation over neutrino energies is concerned, 
``coherent'' refers to a summation in the amplitude whereas ``incoherent''
refers to a summation over squares of amplitudes.}
In the following we adhere to these definitions and
refer the reader to section \ref{discussion} for a clarification of
the notion of neutrino energy in the field-theoretical approach where
the oscillating neutrinos occur in an inner line of the combined
production -- detection Feynman graph.
Having different oscillation lengths in the process under discussion,
then clearly the
relevant coherence length is given by \cite{fukugita}
\be
L^{\mathrm{coh}}_{jk} \equiv \min\limits_a L^{\mathrm{coh}}_{a;\,jk} \,.
\ee
Both kinds of neutrino energy spread, coherent and incoherent, 
lead to a loss of 
the oscillation pattern if $L > L^{\mathrm{coh}}_{jk}$ and 
cannot be distinguished experimentally \cite{KNW96}.
Apart from the condition 
$L \lesssim L^{\mathrm{coh}}_{jk}$ other conditions have to be fulfilled
for the oscillation pattern to be present which do not depend on $L$
\cite{rich,GS96,cam}.

In this paper we will use the LSND experiment \cite{LSND} with the process
\be\label{process}
\mu^+ \to e^+ + \nu_e + \bar\nu_\mu 
\stackrel{\nu\: \mathrm{osc.}}{\leadsto}
\bar\nu_e + p \to n + e^+
\ee
as a model for our
investigation for two reasons: first of all, the $\bar\nu_\mu$
neutrino source ($\mu^+$) is unstable and we want to extend the
field-theoretical approach
of Ref.\cite{GS96} by taking into account the finite
lifetime of the source; secondly, there is a claim made in
Ref.\cite{mohanty} that in the LSND experiment the condition for
coherence is not fulfilled. In the following, we will discuss in
detail the effects of
\begin {enumerate}
\item the quantum-mechanical uncertainties of momentum and energy of the 
initial particles involved in the production and detection  
processes represented by the widths of their respective wave packets or 
stationary states,
\item the finite lifetime of the neutrino source particle,
\item the uncertainties in the measurements of energies and
momenta of the particles in the final state of neutrino production
and, in particular, of the detection process.
\end{enumerate}
We are not able to take into account the interaction of the neutrino source 
particle (the $\mu^+$ in our case) or the interaction of particles 
in the final state of the source process (in our case the positron
originating from the $\mu^+$ decay) with the matter background in
which the source particle is generated (in LSND this background is
water) in the field-theoretical treatment. We will only comment on the
second of these effects in the last section of the paper.
In the wave packet approach it is said that the interaction of the
particles in the source process with the matter background interrupt
neutrino emission and estimates of these effects are used to
determine the ``size of the neutrino wave packet'' \cite{kim}.

To include the finite lifetime of the neutrino source of the process
(\ref{process}) we combine field theory with the Weisskopf--Wigner
approximation \cite{WW} in section \ref{perturbation}. 
In section \ref{amplitude} we calculate the
amplitude for the reaction (\ref{process}) by taking into account that
the distance $L$ between the source and the detector is
macroscopic. This is achieved by using a theorem proved in
Ref.\cite{GS96} and an integral discussed in detail in the appendix of
the present paper. In section \ref{coherence} we derive
conditions for the existence of neutrino oscillations independent of $L$,
while in section \ref{crosss} we study some aspect of the cross
section of the total production -- detection process concerning the
finite lifetime of the source. All the conditions for
neutrino oscillations obtained in sections \ref{coherence} 
and \ref{crosss} -- whether dependent on $L$ or not -- 
are discussed in detail in section \ref{discussion} where
we also study the problem raised in Ref.\cite{mohanty}
and try to elucidate the nature of the neutrino energy spread. 

\section{Perturbation theory and Weisskopf--Wigner approximation}
\label{perturbation}

To fix the notation we shortly repeat the basics of
time-dependent perturbation theory.
We consider a system described by the Hamiltonian
$H=H_0+H_1$
where $H_0$ and $H_1$ are not not explicitly time-dependent. 
The eigenvalues and eigenvectors of $H_0$ will be denoted as in the 
relation $H_0 \phi_j = E_j \phi_j$
where $\{\phi_j\}$ is a complete orthonormal system of states. 
For an arbitrary state $\psi = \sum_j c_j(0) \phi_j$ at $t=0$,
the Schr\"odinger Equation gives the time evolution
$\psi(t) = \sum_j c_j(t) \phi_j e^{-iE_jt}$
where the components $c_j(t)$ obey the relations
\be
i\dot{c}_j(t) = 
\sum_k c_k \langle \phi_j | H_1 \phi_k \rangle e^{i(E_j-E_k)t} =
\sum_k c_k \langle \phi_j | H_{1,\mathrm{int}}(t) \phi_k \rangle 
\ee
and
\be\label{intpict}
H_{1,\mathrm{int}}(t) \equiv e^{iH_0t} H_1(0) e^{-iH_0t}
\ee
defines the interaction Hamiltonian in the interaction picture.

Let us now study the $\mu^+$ decay and the ``subsequent'' 
detection of $\bar\nu_e$ by $\bar\nu_e + p \rightarrow e^+ + n$.
The interaction Hamiltonian is given by
\be
H_1 = H_S^+ + H_S^- + H_D^+ + H_D^-
\ee
where the indices $S$ and $D$ denote source and detection, respectively,
and $H_S^+$ and $H_D^+$ are given by the Hamiltonian densities
\ba\label{HSD}
{\cal H}^+_S & = & \frac{G_F}{\sqrt{2}}  
\bar\mu \gamma_\rho (1-\gamma_5) \nu_S \,
\bar\nu_e \gamma^\rho (1-\gamma_5)e \,, \no \\
{\cal H}^+_D & = & \frac{G_F}{\sqrt{2}} \cos \vartheta_C 
\bar\nu_D \gamma_\lambda (1-\gamma_5) e \,
\bar n \gamma^\lambda (1-g_A \gamma_5) p 
\ea
describing muon and neutron decay, respecitively, and
\be
\nu_S \equiv U_{\mu j}\nu_j\; ,\quad \nu_D\equiv U_{ej}\nu_j\; .
\ee
Actually, $\bar\nu_e$ in Eq.(\ref{HSD}) should be replaced by 
$U^*_{ej}\bar\nu_j$, however, this has no effect on the final result
for neutrinos much lighter than the mass of the muon. The
Hamiltonians with the superscript $+$ are the Hermitian conjugates of
those which carry the minus sign.

Let us now sketch how to incorporate the finite muon lifetime in
perturbation theory \cite{WW}. To this end we take into account $H_D$
only when it occurs together with $c_i(t)$ (see the initial conditions
(\ref{initial})) but take $H_S$ in all
instances. With this proviso the following states and coefficients 
are involved in perturbation theory:
\be\label{states}
\begin{array}{lrlcll}
\mbox{initial state} & 
\mu^+; & p & \leftrightarrow & c_i, & \phi_i \\[0.4mm]
\mbox{intermediate states}\quad &
e^+, \nu_e, \bar\nu_S; & p & \leftrightarrow & c'_j, & \phi'_j \\[0.4mm]
& \mu^+; & e^+, n, \nu_D & \leftrightarrow & c''_k, & \phi''_k \\[0.4mm]
& e^+, \nu_e, \bar\nu_S; & e^+, n, \nu_D & \leftrightarrow &
c'''_{j\otimes k}, & \phi'''_{j\otimes k} \\[0.4mm]
\mbox{final state} &
e^+, \nu_e; & e^+, n & \leftrightarrow & c_f, & \phi_f
\end{array}
\ee
The initial conditions for the coefficients are given by
\be\label{initial}
c_i(0)=1 \quad \mbox{and} \quad 
c_f(0) = c'_j(0) = c''_k(0) = c'''_{j\otimes k}(0)=0 \,. 
\ee
The differential equations for 
the coefficients are
\ba
\label{c}
i\dot{c}_i(t) & \simeq &
\sum_j c'_j(t) \langle \phi_i | H^-_{S,\mathrm{int}}(t) \phi'_j \rangle
\,,\\
\label{cf}
i\dot{c}_f(t) & = &
\sum_j c'_j(t) \langle \phi_f | H^+_{D,\mathrm{int}}(t)
\phi'_j \rangle
+
\sum_k c''_k(t) \langle \phi_f | H^+_{S,\mathrm{int}}(t)
\phi''_k \rangle \,,\\
\label{c'}
i\dot{c}'_j(t) & \simeq &
c_i(t) \langle \phi'_j | H^+_{S,\mathrm{int}}(t) \phi_i
\rangle \,,\\
\label{c''}
i\dot{c}''_k(t) & \simeq &
c_i(t) \langle \phi''_k | H^+_{D,\mathrm{int}}(t) \phi_i \rangle
+
\sum_j c'''_{j\otimes k}(t) \langle \phi''_k | H^-_{S,\mathrm{int}}(t)
\phi'''_{j\otimes k} \rangle \,,\\
\label{c'''}
i\dot{c}'''_{j\otimes k}(t) & \simeq &
c''_k(t) \langle \phi'''_{j\otimes k} | H^+_{S,\mathrm{int}}(t)
\phi''_k \rangle \,.
\ea
With these approximations we get a closed system for $c_i$, $c'_j$,
$c''_k$, $c'''_{j\otimes k}$. 
If we insert Eq.(\ref{c'''}) into Eq.(\ref{c''}) we arrive at
\be\label{c2}
i\dot{c}''_k(t) \simeq c_i(t)
\langle \phi''_k | H^+_{D,\mathrm{int}}(t) \phi_i \rangle -
i\sum_j \langle \phi''_k | H^-_{S,\mathrm{int}}(t) 
\phi'''_{j\otimes k} \rangle \int^t_0 dt' c''_k(t') 
\langle \phi'''_{j\otimes k} | H^+_{S,\mathrm{int}}(t') \phi''_k \rangle \,.
\ee
However, looking at the intermediate states (\ref{states}), we see
that the equations
\be\label{relation}
\langle \phi'''_{j\otimes k} | H^+_{S,\mathrm{int}}(0) 
\phi''_k \rangle =
\langle \phi'_j | H^+_{S,\mathrm{int}}(0) \phi_i \rangle 
\quad \mbox{and} \quad
E''_k - E'''_{j\otimes k} = E_i - E'_j
\ee
hold trivially because 
in the first matrix element $e^+,n,\nu_D$ and in the second
second matrix element the proton are only spectators.
Inserting Eq.(\ref{relation}) into Eq.(\ref{c2}) then with a 
partial integration the second term of Eq.(\ref{c2})
is written as
\be\label{A}
-i \sum_j A_j \left[ \left. 
\frac{e^{i(E_i-E'_j)(t-t')}}{-i(E_i-E'_j)}
c''_k(t') \right|_0^t -
\int_0^t dt' \frac{e^{i(E_i-E'_j)(t-t')}}{-i(E_i-E'_j)} 
\dot{c}''_k(t') \right]
\ee
where
\be
A_j \equiv
| \langle\phi'_j | H^+_{S,\mathrm{int}}(0) \phi_i \rangle |^2 \,.
\ee
We neglect now the term with $\dot{c}''_j$ in
Eq.(\ref{A}) because it is of higher order and 
replace $E_i-E'_j$ by $E_i-E'_j + i\epsilon$
($\epsilon\downarrow 0$) in Eq.(\ref{A})
to have a well-defined expression. Since $c''_k(0)=0$ we obtain
\ba
& & -i\sum_j A_j \left. 
\frac{e^{[i(E_i-E'_j)-\epsilon](t-t')}}{-i(E_i-E'_j) + \epsilon}
c''_k(t')\right|_0^t = 
\sum_j A_j \frac{c''_k(t)}{E_i-E'_j+i\epsilon} \no\\  
& = & \left( \sum_j A_j P \left( \frac{1}{E_i-E'_j} \right) -
i\pi \sum_j A_j \delta(E_i-E'_j) \right)
c''_k(t) = (\Delta E_i-\frac{i}{2} \Gamma) c''_k(t)
\ea
where $P$ denotes the Cauchy's principal value and $\Gamma$ the
total decay width of the muon.
We neglect in the following the energy shift $\Delta E_i$ or we can
think it being already incorporated in the muon mass. Hence we get
\be
i\dot{c}''_k(t) \simeq 
c_i(t) \langle \phi''_k | H^+_{D,\mathrm{int}}(t)\phi_i\rangle 
- \frac{i}{2}\Gamma c''_k(t) \,.
\ee
With similar arguments one obtains \cite{WW}
\be
c_i(t)\simeq e^{-\frac{1}{2}\Gamma t} 
\ee
and therefore
\be
c''_k(t) \simeq -i\int_0^t dt' \langle \phi''_k |
H^+_{D,\mathrm{int}}(t') \phi_i
\rangle e^{-\frac{1}{2}\Gamma t} \,.
\ee
Using Eqs.(\ref{cf}) and (\ref{c'}) we get the final result
\ba\label{CF}
c_f(t) & \simeq & (-i)^2 \int_0^t dt_1 \int_0^{t_1} dt_2 \no\\
&& \times \langle\phi_{fk}|\left(H^+_{D,\mathrm{int}}(t_1)
H^+_{S,\mathrm{int}}(t_2)e^{-\frac{1}{2}
\Gamma t_2}+H^+_{S,\mathrm{int}}(t_1)e^{-\frac{1}{2}\Gamma t_1}
H^+_{D,\mathrm{int}}(t_2)
\right)\phi_i\rangle \,.
\ea
This formula corresponds to the intuitive expectation. Apart from
starting the time integration at the inital time
$t_i=0$ instead of $t_i=-\infty$ we have
the usual time-ordered product with the finite lifetime incorporated
in the exponentials.

\section {The amplitude}
\label{amplitude}

With Eq.(\ref{CF}), the Hamiltonian densities (\ref{HSD})
and Eq.(\ref{intpict}) we can write for the amplitude of 
the process (\ref{process}) in the limit $t \rightarrow \infty$
\ba
{\cal A} & = & (-i)^2 
\langle \nu_e (p'_\nu), e^+_S (p'_{eS}); e^+_D (p'_{eD}), n (p'_n) | \no\\
&& \times T \left[ \int_0^\infty dt_1 \int d^3x_1 
\int_0^\infty dt_2 \int d^3x_2\, {\cal H}^+_{S,\mathrm{int}}(x_1) 
e^{-\frac{1}{2}\Gamma t_1} {\cal H}^+_{D,\mathrm{int}}(x_2) 
\right] |\mu^+ ; p \rangle 
\ea
where $T$ is the time-ordering symbol.
We assume that the muon $\mu^+$ and the proton $p$ are localized at the 
coordinates $\vec x_S$ and $\vec x_D$, respectively. We imagine 
the proton being the nucleus of a hydrogen atom and bound in a molecule.
Therefore we assume the proton state as stationary whereas 
the decaying muon will be described
by a free wave packet with an average momentum equal to zero. This situation
corresponds to the LSND experiment where the $\mu^+$ is assumed to decay at 
rest. Since neutrino production and detection are localized at $\vec x_S$
and $\vec x_D$, respectively, the spinors of the initial particles can be 
written in coordinate space as
\be
\psi_p (x) =  \psi_p (\vec x - \vec x_D) \, e^{-i E_p t}
\ee
and
\be
\psi_\mu (x) = \int \frac{d^3 p}{(2\pi)^{3/2}}\, \widetilde{\psi}_\mu 
(\vec p)\, e^{- i (\vec p \cdot \vec x - E_\mu(\vec p)t)} \times
e^{i \vec p \cdot \vec x_S} \,,
\ee
respectively, with $E_\mu(\vec{p}) = \sqrt{m^2_\mu+\vec{p}\,^2}$. 
The function $\psi_p (\vec y)$ is peaked at $\vec y = \vec 0$
and the wave packet $\widetilde{\psi}_\mu (\vec p)$ 
in momentum space is peaked 
around the average momentum $\langle \vec p\rangle = \vec 0$. 
The final particles will be described by plane waves.

With the neutrino propagators of the mass eigenstate neutrinos
\be
\langle 0 | T [\nu_j(x_1)\bar\nu_j(x_2)] | 0\rangle=i\int
       \frac{d^4q}{(2\pi)^4}\frac{\not\! q+m_j}{q^2-m_j^2+i\epsilon}
       e^{-iq\cdot(x_1-x_2)}
\ee
we obtain the amplitude
\ba
{\cal A} & = & (-i)^2 \frac{G_F ^2 \cos \vartheta_C}{2} \int 
\frac{d^3 p}{(2\pi)^{3/2}} 
\int^\infty _0 dt_1 \int d^3 x_1 \int_0^\infty dt_2 \int d^3 x_2
\int \frac{d^4 q}{(2\pi)^4} e^{-iq \cdot (x_1 - x_2)} \no\\
&& \times 
\exp \{i (p'_\nu + p'_{eS}) \!\cdot\! x_1 + i(p'_n + p'_{eD}) \!\cdot\! x_2\}
\exp \{i (\vec p \!\cdot\! \vec x_1 - E_\mu (\vec p) t_1 - 
\vec p \!\cdot\! \vec x_S)\}
e^{-\frac{1}{2}\Gamma t_1} e^{- i E_p t_2} \no\\
&& \times
\overline{\widetilde{\psi}}_\mu(\vec p) \gamma_\rho 
(1-\gamma_5) i\sum_j U_{\mu j} \,
\frac{\not\! q + m_j}{q^2 - m_j ^2 + i\epsilon}
U^*_{ej} \gamma^\lambda (1- \gamma_5)v_e (p'_{eD}) \no\\
&& \times
J_S^\rho (p'_\nu, p'_{eS}) \,
\bar u_n (p'_n)\gamma_\lambda (1-g_A \gamma_5) \psi_ p (\vec x_2 -
\vec x_D) 
\ea
with
\be
J_S^\rho (p'_\nu, p'_{eS}) =
\bar u_{\nu_e}(p'_\nu)\gamma^\rho (1 - \gamma_5)v_e (p'_{eS}) \,.
\ee
We start with the integration over $t_1$ where we have to calculate the 
integral
\be
\int_0^\infty e^{- i (q_0 - E'_\nu - E'_{eS} + E_\mu (\vec p)) t_1}\,
e^{-\frac{1}{2}\Gamma t_1}\, dt_1 = \frac{1}{i (q_0 - E'_\nu - E'_{eS}
+E_\mu (\vec p)) + \frac{1}{2} \Gamma}\; . 
\ee
For the integration over $t_2$ we use the relation
\be
\lim_{t\rightarrow \infty} \int^t_0 e^{iEt_2}dt_2 = 
i P \left( \frac{1}{E} \right) + \pi \delta (E) \,.
\ee
Hence a factor
\be
iP\left(\frac {1}{q_0 +E'_n + E'_{eD}- E_p}\right) + 
\pi \delta (q_0 + E'_n + E'_{eD}-E_p)
\ee
appears in the amplitude. Furthermore, in the integration over
$\vec x_2$ we use the relation
\be
(2\pi)^{-3/2}
\int d^3x \, e^{-i\vec{k}\cdot\vec{x}}f(\vec{x}+\vec{b}) = 
e^{i\vec{k}\cdot\vec{b}} \tilde{f}(\vec{k})
\ee
where $\tilde{f}$ is the Fourier transform of $f$.
The integration over $\vec x_1$ is again trivial leading to the 
delta function
\be\label{delta}
(2\pi)^3 \delta (\vec q - \vec p\,'\!\!_\nu - \vec p\,'\!\!_{eS} + \vec p)
\,. 
\ee
Thus we obtain
\ba
{\cal A} & = & - \frac{G_F^2 \cos \vartheta_C}{2} i \sum_j
\int d^3 p \int \frac{d^4 q}{(2\pi)^4} 
\exp \{-i \vec p \!\cdot\! \vec x_S - i
(\vec p\,'\!\!_n + \vec p\,'\!\!_{eD}+\vec q) \!\cdot\! \vec x_D \}\no\\ 
&& \times (2\pi)^3
\delta (\vec q - \vec p\,'\!\!_\nu-\vec p\,'\!\!_{eS} + \vec p)
\frac {1}{i (q_0 + E_S) + \frac{1}{2}\Gamma} \left[ iP\left( \frac {1}
{q_0 + E_D}\right)+ \pi \delta (q_0 + E_D)\right]\no\\ 
&& \times \overline{\widetilde{\psi}}_\mu (\vec p)\gamma_\rho (1-\gamma_5)
U_{\mu j} \frac{\not{\!q} + m_j}
{q^2 - m_j ^2 + i\epsilon} U^*_{ej} \gamma^\lambda
(1-\gamma_5)v_e (p'_{eD}) \no\\
&& \times J_S^\rho (p'_\nu, p'_{eS})\, 
\bar u_n (p'_n)\gamma_\lambda (1 -g_A \gamma_5)
\widetilde{\psi}_p (\vec q + \vec p\,'\!\!_n + \vec p\,'\!\!_{eD})
\ea
where we have defined
\be\label{EDES}
E_S \equiv E_\mu(\vec{p}) - E'_\nu - E'_{eS} \quad \mbox{and} \quad
E_D \equiv E'_n + E'_{eD} - E_p \,.
\ee
The integration over $\vec p$ can easily be carried out because of the delta 
function (\ref{delta}) and leads to the amplitude
\ba
{\cal A} & = & -\frac{G^2_F \cos \vartheta_C}{2} 
e^{-i \vec p_1 \cdot \vec x_S -i \vec p_2 \cdot \vec x_D} i \sum_j
\int \frac{d^4 q}{2\pi}  e^{- i \vec q \cdot \vec L}\no\\ 
&& \times \frac {1}{i (q_0 + E_S) + \frac{1}{2}
\Gamma} \left[\pi\delta (q_0 + E_D) + i P\left(\frac{1}{q_0 +E_D}
\right)\right] \no\\
&& \times \overline{\widetilde{\psi}}_\mu (\vec p_1 - \vec q) \gamma_\rho
(1-\gamma_5) U_{\mu j} 
\frac{\not{\!q} +m_j}{q^2 - m_j^2 + i\epsilon} U^*_{ej} \gamma^\lambda
(1-\gamma_5)v_e (p'_{eD}) \no\\ 
&& \times J_S^\rho (p'_\nu, p'_{eS}) \, 
\bar u (p'_n) \gamma_\lambda (1-g_A \gamma_5)
\widetilde{\psi}_p(\vec q + \vec p_2) \label{ampl}
\ea
where
\be
\vec p_1 \equiv \vec p\,'\!\!_\nu + \vec p\,'\!\!_{eS} \,,\quad
\vec p_2 \equiv \vec p\,'\!\!_n + \vec p\,'\!\!_{eD} 
\quad \mbox{and} \quad  
\vec L   \equiv \vec x_D - \vec x_S \,.
\ee
Note that as a consequence of the integration over $\vec p$ we have
$E_S = E_\mu(-\vec{q}+\vec{p}_1) - E'_\nu - E'_{eS}$, i.e., $E_S$ is
now a function of $\vec{q}$.

Now only the integration over $q$ remains. Since we have a delta function
of $q_0$ within the brackets, the integration of the first of the 
two terms of
the amplitude is trivial. We will show in the appendix that the $q_0 $
integration in the second term, which contains Cauchy's principal value,
leads in the limit of a macroscopic distance
$L$ to the same result. In other words, in the limit of macroscopic $L$
we have simply $2\pi \delta(q_0+E_D)$ from the $t_2$ integration.
In this limit we can apply a theorem proved in Ref.\cite{GS96} 
to perform the $d^3 q$
integration and calculate the leading term of the amplitude for large $L$:
\ba\label{ampinfty}
{\cal A}^\infty & = & \sum_j U_{\mu j}U^*_{ej} e^{i q_j L} 
{\cal A}^\infty_j \no \\
& = & \frac{G^2_F \cos \vartheta_C}{2} 
\, \frac{2\pi^2}{L} \,
i \sum_j U_{\mu j}U^*_{ej}  e^{i q_j L}  \frac {1}{i (E_{Sj} -E_D) + 
\frac{1}{2}\Gamma} \no\\
&& \times \overline{\widetilde{\psi}}_\mu (\vec p_1 + q_j \vec l\,) 
\gamma_\rho (1-\gamma_5) (-\!\!\not{k}_j +m_j)
\gamma_\lambda (1-\gamma_5) v_e (p'_{eD}) \no \\
&& \times J_S^\rho (p'_\nu , p'_{eS})
\bar u (p'_n) \gamma^\lambda (1-g_A \gamma_5)
\widetilde{\psi}_p(-q_j \vec l  + \vec p_2)
\ea
where the definition of ${\cal A}^\infty_j$ is obvious, 
$\vec l$ is the unit vector pointing from the neutrino source to the
detection point, $k_j$ are the momenta of the intermediate neutrinos and
\be\label{neu4mom}
k_j \equiv {E_D \choose q_j \vec l} \, ,\quad q_j \equiv 
\sqrt{E_D^2 - m_j^2}\,.
\ee
Note that 
\be\label{ES}
E_{Sj} = E_\mu (q_j \vec l + \vec p_1) - E'_\nu - E'_{eS}
\ee
because by virtue of the theorem in Ref.\cite{GS96} for each $j$ the
vector $\vec{q}$ has to be replaced by $-q_j \vec{l}$.
The irrelevant phase factor occurring in the first line of Eq.(\ref{ampl})
has been dropped in ${\cal A}^\infty$.

\section{Coherence conditions independent of $L$}
\label{coherence}

Inspecting Eq.(\ref{ampinfty}) it is evident that oscillations
involving $m^2_j-m^2_k$ can only take place if \cite{rich,GS96}
\be\label{ACC}
|q_j-q_k| \lesssim \sigma_S \quad \mbox{and} \quad
|q_j-q_k| \lesssim \sigma_D 
\ee
where $\sigma_S$ and $\sigma_D$ are the widths of
$\widetilde{\psi}_\mu$ and $\widetilde{\psi}_p$, respectively.
We call conditions (\ref{ACC}) amplitude coherence conditions (ACC).
If they  are not fulfilled, either by the source wave packet
or the detector wave packet, then 
${\cal A}^\infty_j \times {\cal A}^\infty_k \approx 0$ $(j \neq k)$
which means that the term labelled by $jk$ is suppressed in Eq.(\ref{P}).
In the ultrarelativistic limit Eq.(\ref{ACC}) is rephrased as
\be\label{ACC1}
\frac{\Delta m^2_{jk}}{2E_D} \lesssim \sigma_{S,D} \,.
\ee
Denoting by $\sigma_{xS,\, xD}$ the widths of the wave functions $\psi_\mu(x)$
and $\psi_p(x)$ in coordinate space, respectively, 
then, with the uncertainty relations
$\sigma_{xS,\, xD}\,\sigma_{S,D} \geq 1/2$, Eq.(\ref{ACC1}) is rewritten as
\be\label{ACC2}
\sigma_{xS,\, xD} \lesssim \frac{1}{4\pi} L^{\mathrm{osc}}_{jk} 
\ee
where we have made the identification $E_\nu = E_D$ 
(see Eqs.(\ref{ampinfty}) and (\ref{neu4mom})).

The amplitude ${\cal A}^\infty_j$ contains the factor
\be\label{factor}
\frac{1}{i(E_{Sj}-E_D) + \frac{1}{2} \Gamma}
\ee
which leads to a condition analogous to Eq.(\ref{ACC}) for
neutrino oscillations to take place:
\be\label{SFC}
|E_{Sj}-E_{Sk}| \lesssim \frac{1}{2} \Gamma \,.
\ee
In the following this condition will be called source wave packet --
finite lifetime condition (SFC). In the ultrarelativistic limit
and with $\sigma_S \ll m_\mu$ we obtain
\be
E_{Sj}-E_{Sk} \approx -\frac{\Delta m^2_{jk}}{2 m_\mu E_D} 
(E_D + \vec{l} \!\cdot\! \vec{p}_1)
\ee
and, assuming that ACC holds, we observe that 
$|E_D \vec{l} + \vec{p}_1| \lesssim \sigma_S$
is valid (see the argument of $\widetilde{\psi}_\mu$ in 
${\cal A}^\infty$ (\ref{ampinfty}))
and therefore Eq.(\ref{SFC}) is rewritten as
\be\label{SFC1}
\frac{\Delta m^2_{jk} \sigma_S}{m_\mu E_D} \lesssim \Gamma \,.
\ee
Defining $\Delta v_\mu \equiv \sigma_S/m_\mu$
as a measure for the spread in velocity of the muon wave packet 
and with the muon lifetime $\tau_\mu = 1/\Gamma$ we can 
interpret Eq.(\ref{SFC1}) as
\be\label{SFC2}
\Delta v_\mu \tau_\mu \lesssim \frac{1}{4\pi} L^{\mathrm{osc}}_{jk} \,.
\ee

\section{The coherence length due to the finite $\mu$ lifetime}
\label{crosss}

Having performed all the integrations in the amplitude in the limit 
$L \to \infty$, we will discuss some aspect of the integration
in the cross section. There we have
integrations of the form $d^3 p'/2E'$ for each particle in the final state,
i.e., in our case $\nu_e$ and $e^+_S$ in the source process and 
$n$ and $e^+_D$ in the detector process. In general these integrations
cannot be performed without knowledge of the source and detector wave 
functions. However, for
\be\label{IC}
\Gamma \ll \sigma_{S,D} 
\ee
the factors 
\be\label{factor1}
\left\{(i(E_{Sj}-E_D)+\Gamma/2) \times (-i(E_{Sk}-E_D)+\Gamma/2)\right\}^{-1}
\ee
in the cross section can be used to apply Cauchy's Theorem in
order to obtain the coherence length associated with the finite
muon lifetime. We assume that the ACC and SFC are valid and 
integrating over momenta of the final state of the detector 
leads to an integration in the variable $E_D$ over a particular interval 
containing $E_{Sj}$ $\forall j$
such that the length $\Delta E_D$ of this interval fulfills
$\Gamma \ll \Delta E_D \ll \sigma_{S,D}$. This allows to write
$E_D = \bar{E}_D + \varepsilon$ where $\bar{E}_D$ denotes the
central value of the interval which we define as
\be\label{dED}
\bar{E}_D \equiv \langle E_{Sj} \rangle
\ee
where $\langle E_{Sj} \rangle$ denotes the mean value of the $E_{Sj}$
and the integration variable $\varepsilon$ varies over
$|\varepsilon| \leq \Delta E_D/2$. From the SFC it follows that
$\bar{E}_D \approx E_{Sj}$ $\forall j$ which is exact up to
terms of order $\Gamma$.
With $\Delta m^2_{jk} > 0$ and 
$q_j-q_k \approx -\Delta m^2_{jk}/2E_D$ we observe that the $\varepsilon$ 
integration over the interval on the real axis can be closed via a half-circle
in the upper half-plane since the radius of the half-circle is much larger
than $\Gamma$ and therefore the factors (\ref{factor1}) make the part of the
path in the upper half-plane negligible in the integral.
Because of Eq.(\ref{IC}) and $\Delta E_D \ll \sigma_{S,D}$
this integration does not affect the
source and detector wave functions to a good approximation. 
Then Cauchy's Theorem states that the 
result of the integration is given by making the replacement
\be
E_D \to E_D^{(0)} = \bar{E}_D + \frac{i}{2} \Gamma 
\ee
in the absolute square of ${\cal A}^\infty$ (\ref{ampinfty}).
Inserting $E_D^{(0)}$ into the exponential $-i\Delta m^2_{jk}L/2E_D$
we see that the cross section contains the damping factor
\be\label{damping}
\exp \left( - \frac{\Delta m^2_{jk}\Gamma}{4\bar{E}^2_D}\,L \right) \,.
\ee
As we will discuss in the next section the requirement (\ref{IC}) is
very likely to be fulfilled for LSND with the decay width of the
muon being $\Gamma \approx 3 \times 10^{-16}$ MeV. $\bar{E}_D$
can still be thought of as being identical with $E_D$
(\ref{EDES}) for all practical purposes because this $\Gamma$ is
so small that $\Delta E_D$ can be chosen smaller than any
achievable accuracy for $E_D$ in a real experiment.

\section{Discussion}
\label{discussion}

\paragraph*{The characteristics of the field-theoretical approach:}
In this paper we have used the field-theoretical approach to
discuss neutrino oscillations as we have done in
Ref.\cite{GS96}. In this approach the whole process of neutrino production and
detection is represented by a single Feynman graph such that
the oscillating neutrinos are associated with the \emph{inner}
line of the graph. However, due to the macroscopic distance
between neutrino source and detector this inner line is on-shell
(see Eq.(\ref{ampinfty})) for each neutrino with definite mass
according to a theorem proven in Ref.\cite{GS96}. 
In the present paper we have incorporated the finite lifetime of
the neutrino source which is given in our concrete example of the LSND
experiment (see Eq.(\ref{process})) by
a positively charged muon whose decay is responsible for 
the $\bar\nu_\mu$ neutrinos with which the experiment is performed.
The finite lifetime of the neutrino source prevented us from
using ordinary perturbation theory with an initial time 
$t_i = -\infty$. Instead we took advantage of the
Weisskopf--Wigner approximation which allowed us to take $t_i = 0$  
and to combine in this way the finite muon lifetime with
perturbation theory. 

With the help of the
above-mentioned theorem all integrations in the amplitude of the
combined production -- detection process could be performed
in the asymptotic limit $L \to \infty$.
From the requirement ${\cal A}^\infty_j \approx {\cal A}^\infty_k$ 
(\ref{ampinfty}) we derived 
the amplitude coherence conditions (ACC) (\ref{ACC}), (\ref{ACC1}) 
and the source wave packet -- finite liftime condition (SFC) (\ref{SFC}),
(\ref{SFC1}). For a given mass-squared difference $\Delta m^2_{jk}$
these three conditions are the prerequisites for neutrino
oscillations. If they are not fulfilled, the term with
$\exp (-i\Delta m^2_{jk} L/2E_\nu)$ is suppressed in the oscillation 
probability which means that no neutrino oscillations 
with respect to $\Delta m^2_{jk}$ are possible. 
Here we have identified the neutrino
energy $E_\nu$ with $E_D$ (\ref{EDES}) which is justified in view of the
definition of $q_j$ (\ref{neu4mom})
occurring in the exponentials $e^{iq_j L}$ of
${\cal A}^\infty$ (\ref{ampinfty}). ACC and SFC are both independent of $L$,
therefore no coherence lengths are associated with them.
In coordinate space Eq.(\ref{ACC}) simply means
that the oscillation length must be larger than
the widths of the production and detection wave functions 
(see Eq.(\ref{ACC2})).
The ACC were among the main results of
Refs.\cite{rich} and \cite{GS96} and an analogous condition in
the framework of the wave packet approach was recently emphasized
in Ref.\cite{GK97}. The SFC (\ref{SFC}) says 
that the spreading of the muon wave function\footnote{The muon
wave function is non-stationary.} during the muon lifetime 
should be less than the oscillation length 
in order not to wash out neutrino oscillations 
(see Eq.(\ref{SFC2})). In Ref.\cite{GS96} this condition was not found because
it was assumed that the source wave function is stationary. Clearly, in such 
a case the energy of the neutrino source is fixed and does not depend on $m_j$
and, consequently, no SFC is present.

\paragraph*{Discussion of the LSND experiment:}
Coming back to the LSND experiment,
$\sigma_S$ represents the momentum spread of the stopped muon
and an estimate of it is given by $\sigma_S \lesssim 0.01$ MeV \cite{louis}.
For the detector proton bound in CH$_2$ groups (mineral oil) 
\cite{LSND} it
is reasonable to assume that in coordinate space its wave
function has a spread of around 1 {\AA} and consequently 
$\sigma_D \sim 2 \times 10^{-3}$ MeV.
Dropping now the indices of the mass-squared difference,
with representative values 
$\Delta m^2 \sim 1$ eV$^2$ and $E_\nu \sim 30$ MeV we obtain
$\Delta m^2/2E_\nu \sim 10^{-14}$ MeV and we conclude that the
amplitude coherence conditions are very well fulfilled in the
LSND experiment. Performing an analogous estimate with the SFC we get
$\Delta m^2 \sigma_S/m_\mu E_\nu \lesssim 3 \times 10^{-18} \; \mbox{MeV}
\: \ll \Gamma \approx 3 \times 10^{-16}$ MeV
with the numbers we used before for the ACC. 
From this result we conclude that also the SFC is valid in the context
of the LSND experiment, though, surprisingly, the margin is only given
by two orders of magnitude. Note that for the ACC this margin is
eleven orders of magnitude.

In Ref.\cite{mohanty} it was found that 
$\Delta m^2/E_\nu \lesssim \Gamma$ should hold for coherent neutrino
oscillations. This condition is not fulfilled for LSND.
In this paper with the field-theoretical treatment
we could not recover this condition which looks like the first 
ACC (\ref{ACC1}) with $\sigma_S$ replaced by $\Gamma$.
In this context we want to stress the following: 
in the process (\ref{process}) there are
three different lengths, namely the sizes of the wave packets of
the neutrino source of order $1/\sigma_S$ and of the detector
particle of order $1/\sigma_D$, where $\sigma_S$ and $\sigma_D$
are the widths of $\widetilde{\psi}_\mu$ and $\widetilde{\psi}_p$,
respectively, and
the length $1/\Gamma$ associated with the finite lifetime of the
source. Each of these lengths is uniquely defined
and the function of each of them uniquely
determined in the field-theoretical framework.
However, in the wave packet approach
the distinction between $\sigma_S$, $\sigma_D$ and $\Gamma$ is
not so clear and each of the three lengths could be associated
with the size of the neutrino wave packet and possibly lead to
erroneous conclusions.

\paragraph*{The characteristics of neutrino oscillations:}
We want to emphasize that in the
field-theoretical approach the notion of a neutrino wave packet
does not exist and the questions whether the
neutrino energy or neutrino momentum is fixed or both can vary
do only make sense in connection with the processes of production and
detection. This is because 
only parameters associated with particles of
the exterior legs of the Feyman graph, i.e, with those particles
which are manipulated in the experiment, determine the neutrino
oscillations. Let us notice that, for fixed momenta of the final state
particles of the production and the detection processes,
with Eq.(\ref{ampinfty}) the oscillation probability has the form
\be\label{prob}
P_{\bar\nu_\mu \to \bar\nu_e}(L) \propto \left| \sum_j
{\cal A}^\infty_j U_{\mu j}U^*_{ej} e^{i q_j L} \right|^2
\ee
and we can imagine that the phase factors $e^{i q_j L}$ represent
the plane waves of the different neutrinos mass eigenstates. 
We will use this fact to compare the wave packet approach with the result
of this paper. We arrive at the following characteristics of 
neutrino oscillations in our field-theoretical approach:
\begin{enumerate}
\item 
We have chosen the detector wave function, i.e., the wave function of the 
proton, to be a bound state and therefore the detector wave function does not 
spread in time. This looks physically very reasonable to us 
meaning that the detector is always on and waiting
(see, however, Ref.\cite{KW97} for a discussion of source and detector with
a temporal resolution). 
As a consequence we have $E_\nu = E_D$ 
(see Eqs.(\ref{EDES}), (\ref{ampinfty}) and (\ref{neu4mom})) 
and Eq.(\ref{prob}) suggests that neutrino oscillations take place between
neutrinos with the same energy but with different momenta $q_j$ \cite{GS96}.
\item
The identification $E_\nu = E_D$
allows to determine the neutrino energy with arbitrary 
accuracy by measuring the energies of the 
particles in the final state of the detector 
process, in our case the neutron and the detector positron, 
with arbitrary accuracy. Therefore, one can limit the averaging
over the neutrino energy to an arbitrarily small interval 
-- of course, in practice at the expense of the number of events -- and 
make any coherence length arbitrarily long by performing only detector
manipulations (see Eq.(\ref{Lcoh})). This so-called
restauration of coherence is trivial in our approach.
It agrees with observations in the wave packet treatment
\cite{KNW96,GKL97}. We find no upper limit to the coherence lengths in
contrast to Ref.\cite{GK97} which is due to the fact that we
assume that the detector is sensitive to energies and momenta of the
particles produced in the detector process whereas in
Ref.\cite{GK97} it is assumed that the detector 
measures \emph{neutrino wave packets} of a certain width.
\item
Taking the field-theoretical approach to neutrino oscillations seriously,
assuming that the detector particle is initially in a stationary state
and that the observation of particles associated with the neutrino detection
is done by energy and momentum measurements,
we come to inevitable conclusion that there are no neutrino wave packets
in neutrino oscillations because all summations over neutrino energies 
happen in the cross section and are thus incoherent
summations.\footnote{Though in this paper we consider a neutrino
source at rest, this conclusion and points 1, 2 are also valid for accelerator
neutrinos because the arguments leading to it depend on the neutrino
detection process and not on the production. Formulated in another
way, it is the \emph{detector} which through
its properties determines the nature of the neutrino energy spread.}
Our assumptions include the cases that some
particles are not detected at all or that cuts in energy and momentum
are made. In addition, any further measurements of observables
commuting with the energy and momentum operators performed by the
detector do not change our conclusion \cite{KNW96}.
We think that our conclusion is correct for realistic
experiments. If one assumes instead that, with respect to the
particles in the final state of the detection process, 
the detector is sensitive 
to wave packets of some form then our conclusion is
not valid. However, we do not know if such a detector exists.
\end{enumerate}

\paragraph*{The coherence lengths:}
Let us now assume that the ACC and SFC hold and study the effects of different
energy and/or momentum averaging mechanisms which all lead to
specific coherence lengths. In the light of the above discussion 
all coherence lengths result from incoherent neutrino energy spreads.
There are three types of coherence lengths \cite{KNW96} associated 
with our neutrino production and detection processes (\ref{process}):
\begin{enumerate}
\item[A.] $L^{\mathrm{coh}}_A$ due to the finite muon lifetime,
\item[B.] $L^{\mathrm{coh}}_B$ due to the interruption of the 
neutrino emission because of 
collisions of the source positron with the background and
\item[C.] $L^{\mathrm{coh}}_C$ due to the neutrino energy spread
introduced by the usual imperfect energy measurements of the particles observed
with a realistic detector.
\end{enumerate}

According to Eq.(\ref{damping}) the first of the coherence lengths is
given by
\be\label{LA}
L^{\mathrm{coh}}_A = \frac{4E^2_\nu}{\Delta m^2 \Gamma} \,.
\ee
This coherence length appears if the energy averaging amounts to
a summation of $E_D$ over an interval much larger than $\Gamma$
(see section \ref{crosss}).
With $1/\Gamma$ corresponding to 659 m and 
$\Delta m^2 \sim 1$ eV$^2$, $E_\nu \sim 30$ MeV this coherence
length is around 200 light years and 
completely irrelevant for the LSND experiment\footnote{Note that
even this coherence length could theoretically be overcome by
measuring $E_D$ with a precision better than $\Gamma$ and
chosing events with $E_D$ in a given interval of size 
$\Delta E_D$ smaller than $\Gamma$ (compare section \ref{crosss}
and point 2 in this section).}
and,
therefore, apart from its effect in the SFC, 
the finite lifetime of the muon could have been
neglected as was done in Ref.\cite{GS96} with the lifetime of
the neutrino source nucleus. However, the main point in the
investigation of
$L^{\mathrm{coh}}_A$ was rather to see how it emerges in the
field-theoretical treatment. It is interesting to note that
$L^{\mathrm{coh}}_A$ enters into the cross section through 
$\exp (-L/L^{\mathrm{coh}}_A)$. We have obtained this form of
the damping factor for $\Gamma \ll \sigma_S, \sigma_D$. If this
condition is not fulfilled the damping factor depends on the
form of the functions $\widetilde{\psi}_\mu$, $\widetilde{\psi}_p$. In
the approach using Gaussian wave packets a corresponding damping
factor has the form $\exp \{-(L/L^{\mathrm{coh}})^2\}$
\cite{GK97}, however, it is not obvious how to compare this factor
with the previous one and which coherence length is meant with
$L^{\mathrm{coh}}$. 

$L^{\mathrm{coh}}_B$ is not included in our
treatment because we do not know how to deal with
random collisions of the source positron with the matter
background and the associated heuristic coherence length 
(\ref{Lcoh}) in the
field-theoretical approach. So here we only repeat 
the arguments found in the literature about the neutrino wave packet approach.
There one estimates the mean free path of the
positron from the $\mu^+$ decay in the matter background where
$\mu^+$ is produced. In the LSND experiment 
this background consists of water \cite{LSND}
and according to the rule of thumb presented in
Ref.\cite{kim} this mean free path of the positron is of the
order of centimeters. Then in the heuristic approach 
one would estimate the size $\ell_B$
of the neutrino wave packet to be of the same order of magnitude
\cite{kim}. Thus, adopting the wave packet approach, one would estimate 
$L^{\mathrm{coh}}_B \sim E^2_\nu \ell_B/\Delta m^2$
which is something like four orders of magnitude smaller than 
$L^{\mathrm{coh}}_A$, but still astronomical, 
making the previous consideration of the coherence
length originating from the finite muon lifetime 
(or a wave packet size of order $\ell_A \sim c\tau_\mu = 659$ m) 
spurious.

The coherence length \cite{KNW96}
$
L^{\mathrm{coh}}_C = L^{\mathrm{osc}} E_\nu/ \Delta E_\nu 
$,
where $\Delta E_\nu$ comes from the inability to measure
$E_\nu = E_D$ better than at a certain realistic experimental accuracy, is the
only relevant coherence length in practice \cite{fukugita}. 
In the LSND experiment \cite{LSND}
the detector positron and neutron are
detected in coincidence and, in addition, the energy of the positron
is measured. We guess that $\Delta E_\nu$ is of 
the order of 5 MeV \cite{LSND}, 
thus $L^{\mathrm{coh}}_C$ is probably not more than
several times the oscillation length. This is many orders of magnitude 
smaller than the astronomical coherence lengths $L^{\mathrm{coh}}_{A,B}$.

\paragraph*{Summary:}
Let us summarize the main points.
We have confined ourselves to situations where the neutrino source
is at rest, thus the present investigation is not
straightforwardly applicable to high energy neutrinos (see, however, 
Ref.\cite{cam} for a field-theoretical discussion of accelerator
neutrino experiments). We have assumed that the 
wave function of the detector particle is
stationary. Then the field-theoretical approach to
neutrino oscillations is completely static
and there are no explicit time averages necessary
as in the wave packet approach. With the field-theoretical method
we have clarified, using the model
reaction (\ref{process}), the roles of the
widths $\sigma_S$, $\sigma_D$ and $\Gamma$
in neutrino oscillations where these widths are associated with the
source, the detector and the finite lifetime of the source,
respectively. 
To check the validity of the ACC (\ref{ACC1}) and the SFC (\ref{SFC1})
in a real experiment concrete values of
$\sigma_S$ and $\sigma_D$ have to be chosen. Making a plausible guess for
$\sigma_D$ and using $\sigma_S \lesssim 0.01$ MeV \cite{louis} 
in the case of LSND, the ACC are very well fulfilled and also
the SFC seems to hold safely. 
Finally, if our method
is a correct approach to neutrino oscillations then, in experiments
with the realistic detector properties assumed in this paper, 
oscillations take place between neutrino mass eigenstates with the same
energy but different momenta, there are no neutrino wave packets and
the coherence length in neutrino oscillations results 
from an incoherent neutrino energy spread.

\acknowledgments

W.G. and S.M. thank the organizers of the 5th Workshop on High Energy
Physics Phenomenology, January 12-26, 1998, Pune, India and the Theory Division
of PRL -- Ahmedabad for providing the opportunity to start this
collaboration. S.M. thanks also S. Pakvasa and L.M. Sehgal for useful
discussions. Furthermore, we are indebted to T. Goldman and
W.C. Louis for information on the momentum spread of the stopped muons
in the LSND experiment.

\newpage
\appendix

\section{Integration over $\lowercase{q}_0$}

We consider the integrals
\be\label{int}
I_n = i\int\limits^\infty _ {- \infty} dq_0 \,
\frac {1}{i(q_0 + E_S)+\frac{1}{2} \Gamma}
\,P\left(\frac{1}{q_0 + E_D}\right)\, \frac {(q_0)^n}{q^2_0 - \vec q\,^2 
-m_j ^2 + i\epsilon}
\ee
where $n=0,1$ and
$P$ denotes Cauchy's principal value. We will calculate this integral 
with the help of the residue calculus which gives the formula
\be
\int dx\, P \left( \frac{1}{x-x_0} \right) f(x) = \frac {1}{2} 
\left( \int_{C_R}+\int_{C_L} \right)dx\,\frac{1}{x-x_0}f(x)
\ee
for a function $f$ which is analytic along the real axis. The paths $C_R$
and $C_L$ lead along the real axis except close to $x=x_0$ where
the point $x=x_0$ is circumvented
to its right or to its left in the complex plane, respectively.
In our case $x_0 = -E_D$ and $f$ has three poles of first order at
\be
q^{(1)}_0 = - E_S + \frac{i}{2}\Gamma \,, \quad
q^{(2)}_0 = \sqrt{\vec{q}\,^2 + m_j ^2} -i\epsilon
\quad \mbox{and}\quad
q^{(3)}_0 = -\sqrt{\vec{q}\,^2 + m_j^2}+i\epsilon \,.
\ee
Since we have only one pole below the real axis we close the contour below.
Then we obtain
\be\label{int1}
I_n = -\pi i\left\{
\frac{x_0^n}{(x_0-q^{(1)}_0) (x_0-q^{(2)}_0) (x_0-q^{(3)}_0)}
+
\frac{2(q^{(2)}_0)^n}%
{(q^{(2)}_0-x_0) (q^{(2)}_0-q^{(1)}_0) (q^{(2)}_0-q^{(3)}_0)} 
\right\} \,.
\ee

In the second part of Eq.(\ref{int1}) we use that $E_D > 0$ and perform the
limit $\epsilon \rightarrow 0$ without getting a singular integrand. 
The integral $I_n$ depends on $\vec q$ and
appears in the amplitude in the following way:
\be
{\cal A} = \int d^3 q \, \Phi (\vec q)I(\vec q) e^{-i \vec q \cdot \vec L}
\ee
where $\Phi$ can be read off from Eq.(\ref{ampl}).
For the second term of $I_n$ we can use Lemma 3 of Ref.\cite{GS96} to
show that it decreases like $L^{-2}$ for $L\rightarrow \infty$. Since we
are only interested in the leading term $\propto L^{-1}$
of the amplitude for large $L$
we can neglect the contribution of the second term of $I_n$. It is then easy
to show that the first term in the integral (\ref{int1}) gives exactly
the contribution to the amplitude ${\cal A}^\infty$ (\ref{ampinfty})
as the term with $\pi \delta(q_0+E_D)$.

\end{document}